\documentclass[12pt,a4paper]{article}
\makeatletter
 
 \@addtoreset{equation}{section}
\makeatother
\begin{document}
\newfont{\teneufm}{eufm10}
\newfam\eufmfam
\textfont\eufmfam=\teneufm
\newcommand{\frak}[1]{{\fam\eufmfam\relax#1}}
\font\minmsbm=msbm8 
\font\mineufm=eufm8 
\font\msbm=msbm10 scaled \magstep1
\def\C{\ifmmode{\mbox{\msbm C}}\else{{\msbm C}\ }\fi} 
\def\R{\ifmmode{\mbox{\msbm R}}\else{{\msbm R}\ }\fi}
\def\X{\ifmmode{\mbox{\msbm X}}\else{{\msbm X}\ }\fi}
\def\Z{\ifmmode{\mbox{\msbm Z}}\else{{\msbm Z}\ }\fi}     
\title{String Solitons in M Theory Fivebrane :\\
A possibility of higher order $U(1)$ bundles} 
\author{Kazuyuki FUJII \thanks{e-mail address : fujii@yokohama-cu.ac.jp}\\
Department of Mathematics\\
Yokohama City University\\
Yokohama 236\\
Japan}
\date{}
\maketitle

\baselineskip 17.5pt 
\begin{abstract}
A volume form $H$ on the $n$--dimensional sphere $S^n$ is closed $(dH=0)$, so
that it is locally written as $H=dB$, where $B$ is a $(n-1)$--form.

In the first half we give an explicit form to $B$ and, moreover, a speculation
concerning higher order $U(1)$ bundles.

In the second half we apply our $B$ in the case $n=3$ to a string soliton
solution discussed by Perry and Schwarz.
\end{abstract}
\vskip 5.0mm
\section{Introduction}\label{sect1}
A volume form $H$ of the two dimensional sphere $S^2$ is closed $(dH=0)$ so
that $H$ is locally written as $H=dB$ with some 1--fom $B$.
When $B$ is written out using the cartesian coordinate in $\R^3$, $B$ becomes
a
Dirac monopole field (see Sect.2).
By the way the monopole field is nowadays recognized as a connection of Hopf
$U(1)$ bundle on $S^2$.

In a similar way a volume form $H$ of the $n$--dimensional sphere $S^n$ can be
written as $H=dB$ locally with some $(n-1)$--form $B$.
Here we want to identify $B$ with a higher order connection, $H$ with a higher
order curvature of $B$ likely in the monopole case of Dirac.
For this purpose we must construct a theory of "higher order $U(1)$ bundles".

Higher order connection forms such as 2--form or 3--form gauge fields in the
supergravity theory play an important role in the $M$ theory developing
recently, see, for example, \cite{JP}.
Therefore it is very important for mathematicians to construct such a higher
order theory. Such a theory is now in progress, \cite{AA}.
The aim of this paper is to give a crucial hint to this construction.

On the other hand a non-linear extension of the Maxwell theory was proposed by
Born--Infeld \cite{MB} in 1934.
As the Maxwell theory has a dyon solution in the equations of motion, this
theory has also dyon--like solution,  see Appendix. This theory has not been
paid a sufficient attension to. One of reasons is not easy to make a
quantization. But in the recent development of $M$ theory its importance is
being realized, \cite{MP}, \cite{AAT}.

In \cite{MP} Perry and Schwarz made a theory of interacting self--dual tensor
gauge fields in the six dimensional space--time and showed that it becomes a
Born-infeld like theory after a dimensional reduction to five dimensions. 
Moreover they showed that their theory has a dyon--like solution ( a string
solition solution in their terminology). 
They construct such a solution by making use of volume form of $S^3$, but they
use the Euler coordinate not the cartsian coordinate to express it.
Therefore their method is, in a certain sence, not to fit in with the context.
It seems natural to use the cartesian coordinate not the Euler one.
In fact we reconstruct their solution by using the cartesian one. As a 
bonus the singularity of Dirac string becomes manifest.

The second aim of this paper is to give a mathematical reinforcement to
\cite{MP}.

The contents of this paper are as follows :

\begin{enumerate}
\item Introduction
\item Volume Form on 3--Dimensional Sphere and a Dirac String
\item Volume Form on $n$--Dimensional Sphere and a Dirac String
\item Perry and Schwarz Equation in 6--Dimensional Space--time
\item A String Soliton Solution
\item Multi String Soliton Solutions ?
\item Discussion
\end{enumerate}
\section{Volume Form on 3--Dimensional Sphere and a Dirac String}\label{sect2}

In this section we show our way of thinking and method in detail in the three
dimensional case.

A volume form on the three dimensional sphere $S^3\subset \R^4-\{0\}$ is given
by
\begin{eqnarray}
H 
&\equiv&
\sum^4_{j=1}(-1)^{j-1}\frac{x_j}{r^4}dx_1\wedge\cdots\wedge
\breve{dx_j}\wedge\cdots\wedge dx_4\nonumber\\
&=&
\frac{x_1}{r^4}dx_2\wedge dx_3\wedge dx_4
-\frac{x_2}{r^4}dx_1\wedge dx_3\wedge dx_4\nonumber\\
&{}&+\frac{x_3}{r^4}dx_1\wedge dx_2\wedge dx_4
-\frac{x_4}{r^4}dx_1\wedge dx_2\wedge dx_3\label{s2-1}\\
r^2&=&\sum^4_{j=1}x^2_j.\label{s2-2}
\end{eqnarray}
Then it is easy to see
\begin{equation}
dH=0 \qquad \mbox{on} \qquad \R^4-\{0\},\label{s2-3}
\end{equation}
so that by the Poincare lemma we can write 
\begin{equation}
H=dB \ ; \quad \mbox{$B$ is a 2--form}\label{s2-4}
\end{equation}
locally (which means that $B$ is only defiend on some (small) open set in
$\R^4-\{0\}$).

We look for the explicit form of $B$.
But $H$ is singular at the origin of $\R^4$, so we cannot use the usual method
to construct $B$ (see, for example, \cite{VA}).
We need a little idea.
Making use of the stereographic projection
\[
{\R}^4-\{\mbox{north line of}\ x_4\} \rightarrow {\R}^3,
\]
we have
\begin{equation}
u_j=\frac{x_j}{r-x_4}\quad (j=1 \sim 3)\label{s2-5}
\end{equation}
or, conversely,
\begin{equation}
x_j=r\frac{2u_j}{1+\sum u_j^2}\quad (j=1 \sim 3),\quad 
x_4=r\frac{-1+\sum u_j^2}{1+\sum u_j^2}.\label{s-6}
\end{equation}
Then a little calculation leads to
\begin{eqnarray}
dx_1\wedge dx_2\wedge dx_3&=&(1-\sum u_j^2)(1+\sum u_j^2)^2,\nonumber\\
dx_1\wedge dx_2\wedge dx_4&=&2u_3(1+\sum u_j^2)^2,\nonumber\\ 
dx_1\wedge dx_3\wedge dx_4&=&-2u_2(1+\sum u_j^2)^2,\nonumber\\
dx_2\wedge dx_3\wedge dx_4&=&2u_1(1+\sum u_j^2)^2.\label{s2-7}
\end{eqnarray}
Therefore putting (\ref{s2-5}), (\ref{s2-7}), into (\ref{s2-1}), we have
\begin{equation}
H=\frac{2^3}{(1+u_1^2+u_2^2+u_2^2)^3}du_1\wedge du_2\wedge du_3
\label{s2-8}
\end{equation}
For this $H$ the Poincare lemma can be applied : 
If we set
\begin{equation}
{\tilde B}\equiv\int_0^1\frac{8t^2}{(1+t^2\sum u_j^2)^3}dt\ (u_1du_2\wedge
du_3- u_2du_1\wedge du_3+u_3du_1\wedge du_2),\label{s2-9}
\end{equation}
We have easily 
\begin{equation}
H=d{\tilde B}.\label{s2-10}
\end{equation}
Next we must express ${\tilde B}$ with $x_1\sim x_4$ coordinates.
Since $u_j=\frac{x_j}{r-x_4} (j=1\sim 3)$, we have
\begin{equation}
du_j=\frac
1{r(r-x_4)^2}\{-\sum_{k=1}^4x_j(x_k-r\delta_{k4})dx_k+r(r-x_4)dx_j\},
\label{s2-11}
\end{equation}
to calculate $du_1\wedge du_2$, $du_1\wedge du_3$, $du_2\wedge du_3$ in
${\tilde B}$.
A hard calculation leads to 
\begin{equation}
B=\int_0^1\frac{8t^2}{\{r-x_4+t^2(r+x_4)\}^3}dt\ (x_1dx_2\wedge
dx_3-x_2dx_1\wedge dx_3+x_3dx_1\wedge dx_2).\label{s2-12}
\end{equation}
We leave the proof to the readers. Finally we seek the coefficient of $B$ :%
\begin{equation}
f(r,x_4)\equiv\int_0^1\frac{8t^2}{\{r-x_4+t^2(r+x_4)\}^3}dt.\label{s2-13} 
\end{equation}
For simplicity we set $a=r-x_4$ and  $b=r+x_4$ and calculate $f(r,x_4)$ as
follows.
\begin{equation}
f(r,x_4)=\int_0^1\frac{8t^2}{(a+t^2b)^3}dt
=\frac{\partial}{\partial a}\frac{\partial}{\partial b}\int_0^1\frac
4{a+t^2b}dt
=\frac{\partial}{\partial a}\frac{\partial}{\partial
b}\left\{\frac{4\tan^{-1}\sqrt{\frac ba}}{\sqrt{ab}}\right\}.\label{s2-14}
\end{equation}
Substituting $a=r-x_4$ and  $b=r+x_4$ and calculating (\ref{s2-14}), we
finally
obtain 
\begin{equation}
f(r,x_4)=\frac{\tan^{-1}\left(\sqrt{\displaystyle{\frac{r+x_4}{r-x_4}}}
\right)}{(r^2-x_4^2)^{3/2}}+\frac{x_4}{2r^2(r^2-x_4^2)}.\label{s2-15}
\end{equation}
In summary, we state our conclusion once more :
\begin{eqnarray}
H &=& dB ;\label{s2-16}\\
B &=& f(r,x_4)(x_1dx_2\wedge dx_3-x_2dx_1\wedge dx_3+x_3dx_1\wedge dx_2)
;\label{s2-17}\\
f(r,x_4) &=&
\frac{\tan^{-1}\left(\sqrt{\displaystyle{\frac{r+x_4}{r-x_4}}}\right)}
{(r^2-x_4^2)^{3/2}} + \frac{x_4}{2r^2(r^2-x_4^2)}.\label{s2-18}
\end{eqnarray}
We note that our local 2-form $B$ has a Dirac string.
\section{Volume Form on $n$--Dimensional Sphere and a Dirac
String}\label{sect3}
In this section we give a general formula to $B$ in (\ref{s2-4}) for general
$n$.
A volume form on the $n$-dimensional sphere $S^n\subset\R^{n+1}-\{0\}$ is
given
by
\begin{eqnarray}
H
&=&
\sum^{n+1}_{j=1}(-1)^{j-1}\frac{x_j}{r^{n+1}}dx_1\wedge\cdots\wedge
\breve{dx_j}\wedge\cdots\wedge dx_{n+1},\label{s3-1}\\
r^2 &=& \sum^{n+1}_{j=1}x_j^2.\label{s3-2}
\end{eqnarray}
Since we can check easily 
\begin{equation}
dH=0 \qquad \mbox{on} \qquad \R^{n+1}-\{0\},\label{s3-3}
\end{equation}
we have locally
\begin{equation} 
H=dB \ ;\quad B \mbox{\ is a $(n-1)$--form},\label{s3-4}
\end{equation}
by the Poincare lemma.
To seek this $B$ it seems good to take an analogy of the preceding section.
But  such an easy way is not relevant in this case(calculations like
(\ref{s2-8}) or (\ref{s2-12}) become very  complicated).

Here we make an another approach.
{}From (\ref{s2-18}) we can guess a general form for $B$ : namely,
\begin{equation}
B=\int_0^1\frac{2^nt^{n-1}}{\{r-x_{n+1}+t^2(r+x_{n+1})\}^n}dt\ 
\sum_{j=1}^n(-1)^{n-1+j-1}x_jdx_1\wedge\cdots\wedge
\breve{dx_j}\wedge\cdots\wedge dx_n.\label{s3-5}
\end{equation}
In fact we can show 
\begin{equation}
dB=H.\label{s3-6}
\end{equation}
The proof is not so hard (we leave it to the readers).
Next we calculate
\begin{equation}
f(r,x_{n+1})\equiv\int_0^1\frac{2^nt^{n-1}}{\{r-x_{n+1}+t^2(r+x_{n+1})\}^n}dt.
\label{s3-7}
\end{equation}
For this purpose we must consider two cases according to even $n$ and odd $n$.
For simplicity we set $a=r-x_{n+1}$ and $b=r+x_{n+1}$ and 
\begin{equation}
f(r,x_{n+1})=\int_0^1\frac{2^nt^{n-1}}{(a+t^2b)^n}dt.\label{s3-8}
\end{equation}
This calculation is not so easy.

\underline{$n$ : odd} ($n=2m+1$)
\begin{eqnarray}
f(r,x_{n+1})
&=& 
\frac{2^{2m+1}}{(2m)!}(\frac{\partial}{\partial a})^m(\frac{\partial}{\partial
b})^m\int_0^1\frac 1{a+t^2b}dt\nonumber\\
&=&
\frac{2^{2m+1}}{(2m)!}(\frac{\partial}{\partial a})^m(\frac{\partial}{\partial
b})^m\left\{\frac{\tan^{-1}\sqrt{\frac ba}}{\sqrt{ab}}\right\}.
\label{s3-9}
\end{eqnarray}

\underline{$n$ : even} ($n=2m$)
\begin{eqnarray}
f(r,x_{n+1})
&=& 
\frac{-2^{2m}}{(2m-1)!}(\frac{\partial}{\partial
a})^m(\frac{\partial}{\partial
b})^{m-1}\int_0^1\frac t{a+t^2b}dt\nonumber\\
&=&
\frac{-2^{2m}}{(2m-1)!}(\frac{\partial}{\partial
a})^m(\frac{\partial}{\partial
b})^{m-1}\left\{\frac 1{2b}\log\frac{a+b}a\right\}\nonumber\\
&=&
\frac{-2^{2m-1}}{(2m-1)!}(\frac{\partial}{\partial
a})^{m-1}(\frac{\partial}{\partial b})^{m-1}\left\{\frac
1{a(a+b)}\right\}.\label{s3-10}
\end{eqnarray}
Here in the process of the calculation we used formula 
\begin{eqnarray}
\int_0^1\frac 1{a+t^2b}dt &=& \frac 1{\sqrt{ab}}\tan^{-1}(\sqrt{\frac
ba}),\label{s3-11}\\
\int_0^1\frac t{a+t^2b}dt &=& \frac 1{2b}\log\frac{a+b}a.\label{s3-12}
\end{eqnarray}
We note that our local $(n-1)$--form $B$ has a Dirac string.
A comment is in order. The calculattion of (\ref{s3-9}) is very hard.
This was performed by Suzuki, see \cite{TS}.

A speculation is in order.

In the previous section we set $n=2$ in (\ref{s3-1}).
Then in (\ref{s3-10})
\begin{equation}
f(r,x_3)=\frac 2{a(a+b)}=\frac 1{r(r-x_3)}\label{s3-13}
\end{equation}
and in (\ref{s3-5})
\begin{equation}
B=\frac{-x_1}{r(r-x_3)}dx_2 + \frac{x_2}{r(r-x_3)}dx_1.\label{s3-14}
\end{equation}
If we write $B$ as
\begin{equation}
B\equiv A_1dx_1+A_2dx_2+A_3dx_3\label{s3-15}
\end{equation}
then we have 
\begin{equation}
A_1=\frac{x_2}{r(r-x_3)},\ A_2=\frac{-x_1}{r(r-x_3)},\ A_3=0.\label{s3-16}
\end{equation}
This is just the monopole field of Dirac.
Nowadays, gauge fields of monopole is identified with a connection of Hopf
$U(1)$ bundle on $S^2$ :
\begin{equation}
U(1) \rightarrow S^3 \rightarrow S^2.\label{s3-17}
\end{equation}
For our case we want to identify our $B$ in (\ref{s3-4}) with a higher order
connection, $H (=dB)$ (\ref{s3-4}) with a higher order curvature.
For that purpose we must construct a higher order $U(1)$ bundle on $S^n$ like
the Hopf bundle (\ref{s3-17}).

The mathematical theory for this speculation is still unknown as far as I
know.

Recently in \cite{MI} a discussion of higher order $U(1)$ bundle is made.
But their arguments are  based on higher order de Rham thory (i.e.
$H^{n-1}(X;\C^*)\cong H^n(X;\Z)$ in the integrable case), so that they are not
discussing Dirac strings.

The Dirac string point of view is very important and relevant.

Our local fild $B$ (which is $(n-1)$--form) has a Dirac string.
We believe that our example will give a crucial hint to the construction of
higher order $U(1)$ bundles on some type of manifolds, \cite{AA}.
\section{Perry and Schwarz Equation in 6--Dimensional Space--time}
In this section we give a brief review of Perry--Schwarz theory in the
6--dimensional space--time within our necessity.

We prepre some notations for later convenience.
Let $M^6$ be a 6--dimensional Minkowski space with a metric
$\tilde{\eta}=\mbox{diag}(-1,1,1,1,1,1)$ and $M^5$ be, similarly, a
5--dimensional Minkowski space with a metric $\eta=\mbox{diag}(-1,1,1,1,1)$.

For a two form gauge field $B_{MN}$ on $M^6$, its field strength $H_{MNP}$ is
given
\begin{equation}
H_{MNP}=\partial_MB_{NP}+\partial_NB_{PM}+\partial_PB_{MN}.\label{s4-1}
\end{equation}
It's dual field strength is 
\begin{equation} 
{\tilde H}^{MNP}=\frac{1}{3!}\epsilon^{MNPQRS}H_{QRS}.\label{s4-2}
\end{equation}
We note here our notations :
\begin{equation}
\epsilon^{012345}=-\epsilon_{012345}.\label{s4-3}
\end{equation}
Next we consider the dimensional reduction from $M^6$ to $M^5$ to compare our
theory with that of 5--brane.
Then $B_{MN}$ is decomposed into $B_{\mu\nu}$ and $B_{\mu5}\equiv A_{\mu}$
(where $\mu,\nu \in \{0,1,2,3,4\}$).
For $A_{\mu}$ we set
$F_{\mu\nu}\equiv\partial_{\mu}A_{\nu}-\partial_{\nu}A_{\mu}$ as usual.
Then $H_{MNP}$ is decomposed into
\
\begin{equation}
H_{\mu\nu\lambda}=\partial_{\mu}B_{\nu\lambda}+\partial_{\nu}B_{\lambda\mu}
+\partial_{\lambda}B_{\mu\nu}\label{s4-4}
\end{equation}
and
\begin{equation}
{\cal F}_{\mu\nu}\equiv
H_{\mu\nu5}=F_{\mu\nu}+\partial_5B_{\mu\nu}.\label{s4-5}
\end{equation}
Here we define a dual field
\begin{equation}
{\tilde H}^{\mu\nu}=-\frac
1{3!}\epsilon^{\mu\nu\rho\lambda\sigma}H_{\rho\lambda\sigma}\label{s4-6}
\end{equation}
whose inverse is
\begin{equation}
H_{\mu\nu\lambda}=\frac{1}2\epsilon_{\mu\nu\rho\lambda\sigma}{\tilde
H}^{\rho\sigma}.\label{s4-7}
\end{equation}
We also note that
\begin{equation}
\epsilon^{01234}=-\epsilon_{01234}.\label{s4-8}
\end{equation}
Now we are in a position to state the equation of Perry--Schwarz \cite{MP}:
\begin{equation}
{\tilde H}_{\mu\nu}=\frac{(1-y_1){\cal F}_{\mu\nu}+({\cal
F}^3)_{\mu\nu}}{\sqrt{1-y_1+\frac 12y_1^2-y_2}},\label{s4-9}
\end{equation}
where 
\begin{equation}
y_1\equiv\frac 12\mbox{tr}{\cal F}^2 \quad,\quad y_2\equiv\frac
14\mbox{tr}{\cal F}^4.\label{s4-10}
\end{equation}
Here we note that in (\ref{s4-10}) Lorentz notations in matrix products are
used,
\begin{eqnarray}
\mbox{tr}{\cal F}^2 &\equiv& {\cal F}_{\mu}^{\nu}{\cal F}_{\nu}^{\mu}={\cal
F}_{\mu\nu}{\cal F}^{\nu\mu}=-{\cal F}_{\mu\nu}{\cal
F}^{\mu\nu},\label{s4-11}\\
\mbox{tr}{\cal F}^4 &\equiv& {\cal F}_{\mu}^{\nu}{\cal F}_{\nu}^{\lambda}{\cal
F}_{\lambda}^{\rho}{\cal F}_{\rho}^{\mu}.\label{s4-12}
\end{eqnarray}
A comment about matrix products is in order : 
We have
\begin{equation}
\mbox{tr}{\cal F}^2=\mbox{Tr}({\cal F}\eta)^2,\quad \mbox{tr}{\cal
F}^4=\mbox{Tr}({\cal F}\eta)^4,\label{s4-13}
\end{equation}
where Tr means a usual trace.,$\mbox{Tr}A^2=A_{ij}A_{ji}$ for a square matrix
$A$.
If we remark
\begin{equation}
-\det(\eta+{\cal F}) = \det(E+{\cal F}\eta) = 1-y_1+\frac
12y_1^2-y_2,\label{s4-14}
\end{equation}
(\ref{s4-9}) is also written in matrix form as 
\begin{equation}
{\tilde H}=\sqrt{-\det(\eta+{\cal F})}
\frac{{\cal F}}{1-{\cal F}^2},\label{s4-15}
\end{equation}
see \cite{PSE}.
This form is very clear.

Next let us turn over (\ref{s4-9}) or (\ref{s4-15}) as follows \cite{MP} :
\begin{equation}
{\cal F}_{\mu\nu}=\frac{(1+z_1){\tilde H}_{\mu\nu}+({\tilde
H}^3)_{\mu\nu}}{\sqrt{1+z_1+\frac 12z_1^2-z_2}},\label{s4-16}
\end{equation}
where
\begin{equation}
z_1\equiv\frac 12\mbox{tr}{\tilde H}^2 \quad,\quad z_2\equiv\frac
14\mbox{tr}{\tilde H}^4.\label{s4-17}
\end{equation}
Noting
\begin{equation}
1+z_1+\frac 12z_1^2-z_2=\det(E+i{\tilde H}\eta)=-\det(\eta+i{\tilde
H}),\label{s4-18}
\end{equation}
(\ref{s4-16}) is also written as
\begin{equation}
{\cal F}=\sqrt{-\det(\eta+i{\tilde H})}\frac{{\tilde H}}{E+{\tilde
H}^2}.\label{s4-19}
\end{equation}
Comparing (\ref{s4-15}) with (\ref{s4-19}) we know that this is just a duality
(maybe $T$-duality).
On the other hand the Bianchi identity of $F_{\mu\nu}$ in five dimensins
\begin{equation}
\epsilon^{\mu\nu\rho\lambda\sigma}\partial_{\rho}F_{\mu\nu}=0\label{s4-20}
\end{equation}
and $F_{\mu\nu}={\cal F}_{\mu\nu}-\partial_5B_{\mu\nu}$ in (\ref{s4-5}) lead
to
the action which are looking for 
\begin{equation}
S_6
=
\int d^6x\left\{\frac 12{\tilde H}^{\mu\nu}\partial_5B{\mu\nu}
+
2(\sqrt{-\det(\eta+i{\tilde H})}-1)\right\}.\label{s4-21}
\end{equation}
Namely we have (\ref{s4-20}) as equations of motion of (\ref{s4-21}), which
are
written as (\ref{s4-16}) (or (\ref{s4-19})) using $F_{\mu\nu}={\cal
F}_{\mu\nu}-\partial_5B_{\mu\nu}$ and next turn over to obtain  (\ref{s4-9})
(or (\ref{s4-15})).
This is the main story of Perry--Schwarz \cite{MP}.
The second term in (\ref{s4-21}) is a Born--Infeld like one.
\section{A String Soliton Solution}
In this section we give an another approach to  the construction of string
soliton solution of the Perry--Schwarz equation (\ref{s4-9}).

We want to split $M^6$ into $M^2 \oplus\R^4$ using the metric in \cite{MP}
\begin{equation}
ds^2=-dt^2+(dx_5)^2+\sum_{j=1}^4(dx_j)^2.\label{s5-1}
\end{equation}
The string world is $M^2$ with $\{t,x_5\}$.
We construct a dyon-like solution under this metric.
In \cite{MP} such a solution has been constructed  using the unit sphere in
$\R^4$.
But they use the Euler angles $\theta$, $\phi$ and $\psi$ instead of cartesian
coordinates $\{x_1,\cdots,x_4\}$ to express the unit sphere $S^3 \subset
\R^4$.
They say "the choice of cartesian co-ordinates for the four dimensions
transverse to the string is rather inconvenient for finding solutions to the
field equations" (\cite{MP}, p.14).
But we show that their assertion is not necessarily true.
In fact we construct a dyon--like solution making use of cartesian coordinates
in $\R^4$.
Then as a bonus the singularity of the Dirac string becomes manifest as shown
in section 2.
Let us take the following ansatz :
All the fields are time--independent and moreover, $x_5$--independent.
Under this ansatz the situations become very simple. Then
\begin{equation}
{\cal F}_{\mu\nu}\equiv
F_{\mu\nu}+\partial_5B_{\mu\nu}=F_{\mu\nu}.\label{s5-2}
\end{equation}
Next we take 
\begin{equation}
A_0=\alpha(r),\qquad A_j=0 \quad (j=1\sim 4)\label{s5-3}
\end{equation}
and a Dirac--like monopole field strength $H$ is in a differential form 
\begin{equation}
H=c\sum_{j=1}^4(-1)^j\frac{x_j}{r^4}dx_1\wedge\cdots\wedge
\breve{dx_j}\wedge\cdots\wedge dx_4\label{s5-4}
\end{equation}
just like (\ref{s2-1}). Here $r^2=\sum_{j=1}^4{x_j}^2$ and $c$ is some
constant.
Then components of $H$ are 
\begin{eqnarray}
H_{123}=c\frac{x_4}{r^4} &,& H_{124}=-c\frac{x_3}{r^4},\nonumber\\
H_{134}=c\frac{x_2}{r^4} &,& H_{234}=-c\frac{x_1}{r^4},\quad 
\mbox{and other are zero}.\label{s5-5}
\end{eqnarray}
Then we have easily form (\ref{s4-6})
\begin{equation}
{\tilde H}_{0j}=-\frac{x_j}{r^4} \quad,\quad \mbox{other ${\tilde H}_{ij}$ are
zero},\label{s5-6}
\end{equation}
\begin{equation}
{\cal F}_{0j}=-\alpha'(r)\frac{x_j}r \quad,\quad \mbox{other ${\cal F}_{0j}$
are zero}.\label{s5-7}
\end{equation}
That is to say,
\renewcommand{\arraystretch}{1.8}
\begin{eqnarray*}
{\tilde H}&=&
\left(\begin{array}{ccccc}
0&-c\displaystyle{\frac{x_1}{r^4}}&-c\displaystyle{\frac{x_2}{r^4}}&
-c\displaystyle{\frac{x_3}{r^4}}&-c\displaystyle{\frac{x_4}{r^4}}\\
c\displaystyle{\frac{x_1}{r^4}}&0&0&0&0\\
c\displaystyle{\frac{x_2}{r^4}}&0&0&0&0\\
c\displaystyle{\frac{x_3}{r^4}}&0&0&0&0\\
c\displaystyle{\frac{x_4}{r^4}}&0&0&0&0
\end{array}\right),\\
%
%
{\cal F}&=&
\left(\begin{array}{ccccc}
0&-\alpha'\displaystyle{\frac{x_1}{r^4}}&-\alpha'\displaystyle{\frac{x_2}{
r^4}}&
-\alpha'\displaystyle{\frac{x_3}{r^4}}&-\alpha'\displaystyle{\frac{x_4}
{r^4}}\\
\alpha'\displaystyle{\frac{x_1}{r^4}}&0&0&0&0\\
\alpha'\displaystyle{\frac{x_2}{r^4}}&0&0&0&0\\
\alpha'\displaystyle{\frac{x_3}{r^4}}&0&0&0&0\\
\alpha'\displaystyle{\frac{x_4}{r^4}}&0&0&0&0
\end{array}\right).
\end{eqnarray*}
\renewcommand{\arraystretch}{1.0}
These forms are very clear.
This is the main reason why we used the cartesian coordinates instead of Euler
ones.
Putting these equations into (\ref{s4-9}) and calculating them leads to a
single equation
\begin{equation}
\frac c{r^3}=\frac{\alpha'}{\sqrt{1-\alpha^{'2}}},\label{s5-8}
\end{equation}
or
\begin{equation}
\alpha'=\frac c{\sqrt{c^2+r^6}}.\label{s5-9}
\end{equation}
This is just the equation found in \cite{MP}.
As for the analysis of this equation see \cite{MP}.

On the other hand we have known in Sec.2 2--form gauge fields $B_{\mu\nu}$
giving (\ref{s5-4}) as the field strength.

In summary, we state our result :
The dyon--like solution written in cartessian coordinate is
\begin{equation}
A_0=\alpha(r),\quad A_j=0 \;(1\leq j\leq 4),\label{s5-10}
\end{equation}
\begin{equation}
B_{ij}=cf(r,x_4)\epsilon_{ijk}x_k \quad \mbox{and\ others are zero},
\label{s5-11}
\end{equation}
where $\alpha(r)$ and $f(r,x_4)$ are respectively given by
\begin{eqnarray}
\frac{d\alpha}{dr}&=&\frac c{\sqrt{c^2+r^6}},\label{s5-12}\\
f(r,x_4)&=&\frac
{\tan^{-1}\left(\sqrt{\displaystyle{\frac{r+x_4}{r-x_4}}}\right)}
{(r^2-x_4^2)^{3/2}}+\frac {x_4}{2r^2(r^2-x_4^2)}.\label{s5-13}
\end{eqnarray}
We note that $f(r,x_4)$ has the Dirac string $r-x_4=0$.
\section{Multi String Soliton Solutions ?}
In this section we study wherther the Perry--Schwarz equation (\ref{s4-9})
admits multi string soliton solutions like \cite{PSN} or not.

For the purpose we prepare a harmonic function $\phi$
\begin{equation}
\phi : \R^4\setminus D \rightarrow \R , \quad \Delta\phi=0 \label{s6-1}
\end{equation}
where $D$ is a subset and $\Delta$ is the 4--dimentional Laplacian.
For example
\begin{equation}
\phi(\X)=\sum_{k=1}^m\frac{q_k}{(\X-P^{(k)})^2},\label{s6-2}
\end{equation}
where $D=\{P^{(1)},\cdots,P^{(m)}\} \subset \R^4$ and $q_1,\cdots,q_m$ are
constant.
Hereafter we use $\phi$ in (\ref{s6-2}) for simplicity.
Now let us consider a "volume form" $H$ 
\begin{equation}
H\equiv \sum^4_{j=1}(-1)^{j-1}\partial_j\phi dx_1\wedge\cdots\wedge
\breve{dx_j}\wedge\cdots\wedge dx_4.\label{s6-3}
\end{equation}
Then we have easily
\begin{equation}
dH=\Delta\phi\ dx_1\wedge dx_2\wedge dx_3\wedge dx_4=0.\label{s6-4}
\end{equation}
That is, $H$ is the closed 3--form.
If we set
\begin{equation}
\phi=\frac c2\frac 1{r^2} \label{s6-5}
\end{equation}
where $c$ is constant, then 
\begin{equation}
\partial_j\phi=-c\frac{x_j}{r^4}.\label{s6-6}
\end{equation}
In this case $H$ in (\ref{s6-3}) reduces to $H$ in (\ref{s5-4}).
Therefore $H$ is a much generalization of $H$ in (\ref{s5-4}).
Components of $H$ are
\begin{equation}
H_{ijk}=-\epsilon_{ijkl}\partial_l\phi\label{s6-7}
\end{equation}
and this dual is
\begin{equation}
{\tilde H}_{0j}=\partial_j\phi \quad \mbox{and} \quad {\tilde
H}_{ij}=0.\label{s6-8}
\end{equation}
Next setting
\begin{equation}
A_0=\alpha(\X; P^{(1)},\cdots,P^{(m)}; q_1,\cdots,q_m), \quad \mbox{and} \quad
A_j=0\ (1\leq j\leq 4),\label{s6-9}
\end{equation}
we have 
\begin{equation}
{\cal F}_{0j}=-\partial_j\alpha  \quad \mbox{and} \quad {\cal
F}_{ij}=0.\label{s6-10}
\end{equation}
Putting these equation into (\ref{s4-9}) and calculating them, we have a
single
equation
\begin{equation}
\partial_j\alpha=\frac{-\partial_j\phi}{\sqrt{1+\sum_j(\partial_j\phi)^2}} \
(1\leq j\leq 4).\label{s6-11}
\end{equation}
Let us study this equation.
{}From $\partial_i\partial_j\alpha=\partial_j\partial_i\alpha \ (i\ne j)$ we
have easily
\begin{equation}
0=\partial_i\phi\partial_jT - \partial_j\phi\partial_iT \equiv
\{\phi,T\}_{i,j}
\label{s6-12}
\end{equation}
where $T=\sum_k(\partial_k\phi)^2$. For $\phi$ in (\ref{s6-2}).,
\[
\phi(\X)=\sum_{k=1}^m\frac{q_k}{(\X-P^{(k)})^2},
\]
we have only
\begin{equation}
m=1 \label{s6-13}
\end{equation}
to satisfy (\ref{s6-12}).

We summarize our result again :
Under our ansatz in Sec.5 multi string soliton solutions like (\ref{s6-2})
don't be admitted.
But our ansatz to solve (\ref{s4-9}) is very restrictive.
we don't know until now whether the Perry--Schwarz equation admits multi
string
soliton solutions or not under more general ansatz.
This is an interesting problem.
\section{Discussion}
We discuss in this paper a possibility of theory of " higher order $U(1)$
bundles" in the basis of the volume form on the $n$--dimensional sphere $S^n$.

We also applied our method to reconstruct a string soliton solution discussed
by Perry and Schwarz in the 3--dimensional case $S^3$.

The Perry and Schwarz equation is attractive and rich enough to construct
further solutions and, moreover, their moduli spaces.

On the other hand a non--abelian extension of Born--Infeld theory has been
constructed by Tseytlin \cite{AAT}.
His model becomes the Yang--Mills one in the weak coupling regions.
The  Yang--Mills one has an instanton solution (just like a dyon solution in
the Maxwell theory).

But his model is very complicated so that we cannot write off even its
equations of motion.
We are interested in whether this model has an instanton--like solution or not
(just like a dyon--like solution in the Born--Infeld theory).
As for this point see Fujii \cite{KF}.
Further work on this subject is needed.

\medskip
Acknowledgements\\
The author is very grateful to Akira Asada, Shinichi Nojiri, Kiyotomo Ozawa
and
Tatsuo Suzuki for valuable discussions and to Michiko Kasai for typing this
draft.

He was partially supported by the Grant-in -Aid for Scientific Research
09640210.
\appendix
\section{Appendix}
We review a dyon--like solution of the Born--Infeld theory, see \cite{MB},
\cite{MK}.
Let $M^4$ be a 4--dimentional Minkowski space and $\eta$ a metric.,
$\eta=\mbox{diag}(-1,1,1,1)$.
For gauge fields $A_{\mu} (\mu=0\sim 3)$ on $M^4$ a curvature $F_{\mu\nu}$ is
defined by $F_{\mu\nu}=\partial_{\mu}A_{\nu}-\partial_{\nu}A_{\mu}$.
As usual we set
\begin{eqnarray}
E_j
&\equiv&
F_{0j} \quad (j=1\sim 3)\label{a-1}\\
B_j
&\equiv&
\frac 12\epsilon_{jkl}F^{kl}
\Leftrightarrow
B_1=F_{23},B_2=-F_{13},B_3=F_{12}.\label{a-2}
\end{eqnarray}
For each $j$, $E_j$ (resp. $M_j$) is a electric(resp. magnetric) field
strength.
The Lagrangean of (source--free) Maxwell theory is given by
\begin{equation}
{\cal L}_M\equiv -\frac
12F_{\mu\nu}F^{\mu\nu}=\sum_{j=1}^3(E_j^2-B_j^2).\label{a-3}
\end{equation}
It is well--known that this theory has a dyon solution.

Next we turn to the Born--Infeld theory.
First we prepare the curvature matrix 
\begin{equation}
F \equiv(F_{\mu\nu}) =
\left(\begin{array}{cccc}
0&E_1&E_2&E_3\\
-E_1&0&B_3&-B_2\\
-E_2&-B_3&0&B_1\\
-E_3&B_2&-B_1&0
\end{array}\right).\label{a-4}
\end{equation}
The Lagrangean of Born--Infeld theory is given by
\begin{equation}
{\cal L}_{BI}\equiv\frac 2{g^2}(1-\sqrt{-\det(\eta+gF)}),\label{a-5}
\end{equation}
where $g$ is a coupling constant.
Then it is easy to see 
\begin{equation}
-\det(\eta+gF)=1-g^2\sum_{j=1}^3(E_j^2-B_j^2)-g^4(\sum_{j=1}^3E_jB_j)^2,
\label{a-6}
\end{equation}
so that we have 
\begin{equation}
{\cal L}_M={\cal L}_{BI} \label{a-7}
\end{equation}
in the weak coupling regions $(0<g\ll 1)$.
Threfore ${\cal L}_{BI}$ is a non--linear extension of ${\cal L}_M$.
In the following we get set $g=1$ for simplicity.

Next we look for equations of motion of (\ref{a-5}) and a dyon--like solution.

We set
\begin{equation}
{\tilde G}_{\mu\nu}\equiv-\frac{F_{\mu\nu}-\beta{\tilde
F}_{\mu\nu}}{\sqrt{-\det(\eta+F)}},\label{a-8}
\end{equation}
where
\begin{equation}
\beta=\sum_{j=1}^3B_jE_j \quad \mbox{and} \quad 
{\tilde F}_{\mu\nu}\equiv\frac
12\epsilon_{\mu\nu\lambda\rho}F^{\lambda\rho}.\label{a-9}
\end{equation}
We note here that $\epsilon^{0123}=-1=\epsilon_{0123}$.
Then equations of motion read
\begin{equation}
\partial^{\mu}{\tilde G}_{\mu\nu}=0.\label{a-10}
\end{equation}
To seek a dyon--like solution of (\ref{a-10}), we take an ansatz :
\begin{equation}
A_0=\alpha(r),\quad A_1=\frac{-x_2}{r(r-x_3)},\quad
A_2=\frac{x_1}{r(r-x_3)},\quad A_3=0,\label{a-11}
\end{equation}
where $r^2=\sum_{j=1}^3x_j^2$.
The gauge fields $\{A_1,A_2,A_3\}$ are just monopole ones.
Then we obtain easily from (\ref{a-11}) 
\begin{equation}
E_j=-\alpha'\frac{x_j}r,\quad
B_j=-\frac{x_j}{r^3},\quad (j=1\sim 3).\label{a-12}
\end{equation}
Putting (\ref{a-12}) into (\ref{a-10}) we have a single equation which are
looking for
\begin{equation}
\frac{d\alpha}{dr}=\frac c{\sqrt{r^4+c^2+1}},\label{a-13}
\end{equation}
where $c$ is a integral constant (because (\ref{a-10}) is a second order).
Compare this (\ref{a-13}) with (\ref{s5-12}).
%

%
\end{document}